\documentclass[11pt]{article}

\begin{document}
\begin{center}
\quad\\
{\LARGE Exact solution of an exclusion process with three
 classes of  particles and vacancies}
\quad \\ \quad \\ \quad \\
{\Large K.~Mallick,{\footnote {\tt e-mail: mallick@spht.saclay.cea.fr}} 
S.~Mallick,$^2$ and\setcounter{footnote}{2}
N.~Rajewsky {\footnote {\tt e-mail: rajewsky@math.rutgers.edu}}}
\end{center}
\vskip1cm 
\begin{center}
$^1$\, Department of Physics\\
Technion, 32000 Haifa, Israel and\\
Service de Physique theorique\\ C. E. Saclay, 
F-91191 Gif-sur Yvette Cedex, France\\
\quad \\
${^{2}}$\, Institut d'Optique\\
Centre scientifique  d'Orsay B.P. 147 \\
91403 Orsay Cedex, France\\
\quad \\
${^{3}}$\,
\,Department of Mathematics\\ Rutgers University,
New Brunswick\\ New Jersey 08903, USA.
\end{center}
 
\begin{abstract}
 We present an exact solution for an asymmetric exclusion
 process on a ring  with three classes of particles and vacancies.
 Using a matrix product Ansatz, we
 find explicit expressions for the weights of the 
 configurations in the stationary state. The solution
 involves tensor products of quadratic algebras.
\end{abstract}
 

\section{Introduction}
 The one-dimensional asymmetric simple  exclusion process (ASEP) has been
 extensively studied in mathematical and physical literature
 as one of the simplest models for non-equilibrium statistical
 mechanics \cite{liggett,spohn,zia,derr}. The ASEP is a model of particles
 diffusing  on a lattice  driven by
 an external field and with hard-core exclusion.
 A variety of
 different phenomena can be described  by the exclusion process,
 for instance superionic conductors \cite{rich}, traffic flows \cite{nagel}
 and interface growth \cite{krug}.
 Exact
 results have been obtained for the one-dimensional
 exclusion process  with the help of various methods such as 
 the Bethe Ansatz \cite{dhar,spohn2,derrleb}, and, more recently, 
 a matrix product Ansatz (see for example \cite{derr2}).

 The matrix Ansatz has led to new exact results concerning
 the stationary state of various models.
 Originally developed for the study of directed animals on 
 a lattice \cite{hakim}, this method has been successfully applied
 to the exclusion
 process with open boundaries \cite{DEHP}. It   has then
 been extended to  study  systems with second class particles
 and  shocks \cite{DJLS}, time discrete
 dynamics \cite{hinrich, nikolaus1, nikolaus2, nikolaus3, peschel, evans}, 
 and to calculate 
  diffusion constants
 \cite{DM}. 
 The algebras involved have led  to interesting representation problems
 \cite{rittess}.

  Models with more than two classes  of particles have hardly  been
  investigated  (\cite{foulad}, \cite{martin}). Here,  
  we study  an exclusion model with vacancies and 
  three classes of particles on a ring. Up to now, it was not 
  known whether  the matrix Ansatz could be used 
  to construct an exact solution  of this model.
  In this paper,
  we shall  give an exact expression for the stationary
  state of our model by
  using a suitable matrix Ansatz that 
  involves tensor products  of quadratic algebras. 
  Apart from the question as to 
   how far the matrix Ansatz can be used, the model is interesting
  in  itself, because  it is suitable for a detailed study of shocks
  \cite{prep}. 
  In section 2, we define the model and explain
  what the matrix Ansatz is. In section 3 we recall the matrix
  solution  for the asymmetric exclusion process with second class 
  particles.
  In section 4, we give an explicit representation of  the
  algebra that describes  the stationary state of our 
  model and  present
  a proof of our solution. We also give an explicit solution
  for the stationary state without using any representations.
  We then discuss an algorithm 
  that allows to obtain exact properties
  of the stationary state for large systems 
  by using a  computer. 
  The concluding section discusses our  results and
  some generalizations. The appendix contains  details
  of the proof and certain algebraic properties and
  recursion relations.

 \section{Definition of the model}
   We consider a periodic
 one dimensional lattice of $L$ sites. Each site of the lattice is 
 either empty or 
 occupied by  one  single particle that  can be of  type 1,2 or 3.
 For reasons that will become apparent later on, 
 we say that empty sites are occupied by holes (or vacancies)
 and we shall call   holes as {\it   particles of the fourth  type.}
   We  denote the number
 of particles of type $k$ in the system by $n_k$, where  
 $k=1,2,3,4$. The state of a site $i$ is  
 specified by a discrete variable $\tau_i$
 that takes four possible values:
 \begin{equation}
  \tau_i = 1,2,3 \hbox { or } 4 \hbox { if  site $i$ is
 occupied by a particle of type 1, 2, 3 or 4} \,\,\,\, .
 \end{equation}

  The  dynamics of the system  is given  by certain transition rules. 
  During an infinitesimal time step $dt$, the following processes
  take place on a bond $(i,i+1)$ with probability $dt$:

 \begin{eqnarray}
   12 &\to& 21 \hbox { with rate } 1 \nonumber \\
   13 &\to& 31 \hbox { with rate } 1 \nonumber \\
   14 &\to& 41 \hbox { with rate } 1 \nonumber \\
  \nonumber \\
   23 &\to& 32 \hbox { with rate } 1 \nonumber \\
   24 &\to& 42 \hbox { with rate } 1 \nonumber \\
   \nonumber \\
   34 &\to& 43 \hbox { with rate } 1 \,\,\,\,\,\, .
 \label{rules}
 \end{eqnarray}
  All other transitions are forbidden. Obviously, the dynamics conserves
  the number of particles and  one has $\sum_{k=1}^4n_k=L$.
  It is also clear from these rules that particles
  of type $n$ can `overtake'  particles of type type $m$ only
  if $n<m$. The transition rules therefore induce 
  a hierarchy among the particles. 
  
  In the literature, particles of
   type $k=1,2,3$ are named  {\it first}, {\it second}
   and {\it  third class particles}.
  The model defined in (\ref{rules}) is called
  `the asymmetric exclusion process with three classes of particles
 and holes'.
  
  Note that a first class particle
  behaves always in the same way in regards to all the other
  particles, whereas  a third class particle
   for example 
  behaves like a first class particle in respect to the
  holes, but as a hole in respect to the second and first class
  particles.

 The rules  given in  (\ref{rules})  are  translationally invariant.
 Using this property, we  decide that a particle
 of the third class occupies  the site number $L$
 and we enumerate the different configurations of the system.
 The total number of configurations $N_{tot}$ is  given
 by 
\begin{equation}
 N_{\hbox{tot}} = {(L-1)! \over n_1!n_2!(n_3-1)!n_4! } \ \ .
\end{equation}
  
 The dynamics of the system can be encoded in a Markov 
 matrix $M$ of size $N_{\hbox{tot}} \times  N_{\hbox{tot}}$.
 The coefficient
 $M({\cal C},{\cal C}')$ of this
 matrix represents the rate of transition
 from a  configuration ${\cal C}$ to
 a different configuration ${\cal C}'$;
  $M({\cal C},{\cal C})$ is the rate of exit  from a given
 configuration ${\cal C}$.
 Due to the local structure of the rules (\ref{rules}), $M$ 
 can be written as a sum of
 local operators that represent the transitions
 that take place at  a bond
  $(i,i+1)$
\begin{equation}
 M = \sum_{i=1}^{L} m_{i,i+1} \,\,\,\, .
\end{equation}
 An explicit expression of the matrices $ m_{i,i+1}$ is given in
 the appendix.

  In the long time limit, the process  reaches a stationary state in
 which each configuration ${\cal C}$  of the system has a stationary
 probability $p({\cal C})$. The  stationary state exists and is unique. 
  This  follows from the fact that the rules (\ref{rules})
 define an  irreducible  Markov process, i.e.  any given
 configuration can evolve to any other one. 
 The properties
  of the stationary state can be determined once the probabilities
  $p({\cal C})$ are known for all ${\cal C}$. 
  In  equilibrium statistical mechanics
 these numbers  are given by the Boltzmann factor, but in
 our model there is   a priori no  method to  calculate
 these quantities: one has to solve the stationary master equation
 \begin{equation}
 \sum_{{\cal C}'} M({\cal C},{\cal C}')p({\cal C'})  = 0 \,\,\,\, .
\label{markov}
 \end{equation}
 This is  a system of $N_{\hbox{tot}}$ coupled linear equations  whose
 complexity grows exponentially with the  size $L$ of the system.

  The matrix Ansatz \cite{DEHP}  consists in 
  solving  the system
 (\ref{markov}) by writing the probabilities  $p({\cal C})$ as 
  a trace of a product of four  non-commuting
 operators,  $A_1$, $A_2$, $A_3$ and  $A_4,$ where
 each  represents  one  type of particle:
\begin{eqnarray}
 p({\cal C}) = \frac{1}{Z}{\rm Tr}(A_{\tau_1}...A_{\tau_L})
 \label{ansatz}
\end{eqnarray}
 where $A_{\tau_i}$
 is equal to $A_k$ if  site $i$ is occupied by a particle of type
 $k$ $ (k = 1,2,3,4)$  in  configuration ${\cal C}$.
  The constant $Z$ is an overall normalization factor, that  depends
 on $L$ and on  the $n_k$'s;  it  ensures that $\sum p({\cal C}) = 1$.

 In section 4,  we shall  present explicit operators $A_{k}$ ($k=1,2,3,4$)
 and prove that the weights $ p({\cal C})$  constructed from
 (\ref{ansatz})
 are solutions of  the master equation (\ref{markov}).
 Certain properties of the  solution 
 of the model with just
 first and second class particles (i.e in our case $n_3=0$ )
 are  useful for 
 constructing  the operators $A_k$.
 Therefore,  we   review this
 solution in the following section.

 \section{Matrix solution of  the ASEP  with
 first and   second  class particles}
  Suppose first  that   there are only
  first  class particles and holes on the ring. In this case,
 the steady state  is such that all configurations ${\cal C}$
 have the same weight \cite{ball}.
 Assuming that the last site is always occupied by a 
  first  class particle, we obtain that 
 the total
 number of configurations will be
 $ {(L-1)! \over
 (L-n_1)!(n_1-1)! }$ and each configuration has  a stationary
 probability equal to:
\begin{eqnarray}
 p({\cal C}) = \frac{(n_1-1)!(L-n_1)!}{(L-1)!}
\end{eqnarray}
 In this simple case, a matrix Ansatz is not needed 
  (one  can choose the matrices  representing  particles and holes
  to be both equal to the scalar~1).

  We consider now the model defined in (\ref{rules}) without particles
  of type 3 (i.e. $n_3 = 0$).
   There are $n_1$ particles of the first  
  class and $n_2$  particles
  of the second  class.  For this model, the stationary 
  probability is non-uniform
 and was obtained in \cite{DJLS} 
 from a 
 matrix product Ansatz. Following \cite{DJLS}, we denote  by $D$, $E$
 and $A$ the operators
 that represent particles of type 1, 2 and  holes, respectively.
 The numbers $ p({\cal C})$ obtained  from expression (\ref{ansatz})
 are  the stationary probabilities  of the  exclusion process
 with first and second class particles, 
 if the  three operators $E,$ $D$ and $A$
 generate the  quadratic algebra defined by the relations 
 \begin{eqnarray}
        DE &=& D + E \nonumber \\
        DA &=& A   \nonumber \\
        AE &=& E
 \label{quadr}
 \end{eqnarray}

 It is convenient to work with an explicit representation of the
 algebra (\ref{quadr}). A particularly useful choice is:
 \begin{eqnarray}
  D  = \left( \begin{array}{cccccc}
                  1&1&0&0&. &.\\
                  0&1&1&0&& \\
                  0&0&1&1&&\\
                  0&0&0&1&.&\\
                  . &&&&. &. \\
                  . &&&&&.  
                   \end{array}
                   \right)
 \,\, , \,\, 
  E  = \left( \begin{array}{cccccc}
                  1&0&0&0&&\\
                 1&1&0&0&. &.\\
                 0&1&1&0&& \\
                  0 &0&1&1&&\\
                  . &&&.&. & \\
                  .&&&&.&.  
                   \end{array}
                    \right) \,\; \nonumber \\
   A = |1\rangle\langle{1}| = \left( \begin{array}{cccccc}
                  1&0&0&0&&\\
                 0&0&0&0&. &.\\
                 0&0&0&0&& \\
                  0 &0&0&0&&\\
                  . &&&.&. & \\
                  .&&&&.&.  
                   \end{array}
                    \right) \,\;
 \label{repr}
 \end{eqnarray}

   The operators $D$ and $E$ are represented  by 
  matrices  that act on an infinite dimensional space with denumerable
 basis $(|1\rangle,|2\rangle,...|n\rangle,...)$.
  The operator $A$ is a projector of rank 1
 on the first element of the basis and  has a finite trace. 
 This  ensures that any expression of the type (\ref{ansatz}) is finite.

 Using the algebraic rules (\ref{quadr}) or the explicit
 representation (\ref{repr}),   all stationary probabilities
 are determined.
  In order to calculate  physical quantities
 such as density profiles,
 or average local currents, we  must know   the constant  $Z$, which
 plays a role analogous to that of the  partition function. The  expression
 for   $Z$ is simple
 if there is only one second class  particle in the system,
 always located on the last site.
 In that case, one obtains \cite{DJLS}
 $$ Z = \frac{1}{L} {L  \choose  n_1}{L   \choose  n_1 + 1 }\,\,\,\, .$$
 When the density of second class particles is finite, 
  asymptotic formulae for $Z$ are derived 
 for   systems of large  size, using
 a grand canonical formalism  \cite{DJLS}.

 \section{Matrix solution of the ASEP  with first,
 second  and third class particles}
 \subsection{Explicit forms of the matrices}

 We shall show that the stationary weights,
 solutions of  the master equation (\ref{markov}),
 can be calculated
  via the Ansatz (\ref{ansatz}),  from the following four operators
\begin{eqnarray}
\hspace{-0.in} A_1 &=&  \left( \begin{array}{cccccc}
                  D&0&E&0&0&.\\
                  0&D&0&E&0&. \\
                  0&0&D&0&E&.\\
                  0&0&0&D&0&.\\
                  0&0&0&0&D&. \\
                  . &.&.&.&.&.  
                   \end{array}
                   \hspace{0.2in} \right),
       \nonumber 
 A_2 =  \left( \begin{array}{cccccc}
               D&-E&0&0&0&. \\
               0&0&0&0&0&.  \\
               0&0&0&0&0&.  \\
               0&0&0&0&0&.  \\
               0&0&0&0&0&.  \\
              . &.&.&.&.&. \end{array}
                   \hspace{0.2in} \right)
\hspace{0.2in}\\
A_3 &=&   \left( \begin{array}{cccccc}
               E&0&0&0&0&. \\
               D&0&0&0&0&.  \\
               0&0&0&0&0&.  \\
               0&0&0&0&0&.  \\
               0&0&0&0&0&.  \\
               .&.&.&.&.&. \end{array}
                   \hspace{0.2in} \right),
 A_4 =  \left( \begin{array}{cccccc}
                  E&0&0&0&0&.\\
                  0&E&0&0&0&. \\
                  D&0&E&0&0&. \\
                  0&D&0&E&0&. \\
                  0&0&D&0&E&. \\
                  .&.&.&.&.&.
                   \end{array}
                   \hspace{0.2in} \right).
\label{mat}
\end{eqnarray}
 All matrices are infinite dimensional and 
 their  coefficients  are themselves  the infinite
 dimensional  operators $D$ and $E$ of (\ref{repr})
 which satisfy
 $DE = D + E$ and do not commute with each other (i.e.
  scalar representations of $D$ and $E$ are excluded).
 Another way to look at  the operators
 given in (\ref{mat}) is to consider them as matrices operating
 on an infinite dimensional space, with non-commutative elements.
 The operators $A_2$ and $A_3$ have only two non-zero elements  and 
 the following relation holds:
\begin{eqnarray}
  A_2 A_3 = \left( \begin{array}{cccccc}
                 A&0&0&0&0&.\\
                 0&0&0&0&0&.\\
                 0&0&0&0&0&. \\
                 0&0&0&0&0&. \\
                 0&0&0&0&0&. \\
                 .&.&.&.&.&.  
                   \end{array}
                    \right) \,\,\,\, , 
\label{proj}
\end{eqnarray}
 here $A$ is the rank one projector of  (\ref{repr}). 
 Before we prove that the stationary probabilities
 given in terms of
 the $A_i$'s solve  the master equation
 (\ref{markov}), we have to ensure that they
 are finite. This is not obvious
 because  none of the  operators given in (\ref{mat})
 has  a finite trace.

\subsection{Proof of the finiteness  of the Ansatz}

 We rewrite the expression (\ref{ansatz}) for the
 stationary weights as follows
\begin{eqnarray}
 p({\cal C}) = \frac{1}{Z}{\rm Tr}(A_{\tau_1}...A_{\tau_L})
    =  \frac{1}{Z}{\rm Tr}(YA_{2}XA_{3}T)
\label{facto}
\end{eqnarray}
 where $Y$ and $T$ are products of 
 the  $A_k$'s  $(k=1,2,3,4)$
 and $X$ is a product of $p$ $(p \leq  L-2)$ matrices
 $A_1$ and $A_4$ only. Such a factorization is  possible:
 the term  $A_{2}XA_{3}$  starts
  from  the furthermost (proceeding  from left to right)
 factor  $A_2$ in 
 $(A_{\tau_1}...A_{\tau_L})$ and ends when  an
 $A_3$  appears for the first time after this  $A_2$.
  Such a  factor $A_3$ always exists  since  $A_{\tau_L} = A_3 .$

 We now prove that the matrix $A_{2}XA_{3}$, where
 $X$  is a product of $p$ factors  $A_1$ and $A_4$, 
  can have non-zero elements
 only in its first $(p+2)$ lines or columns. We shall
 say, in such a case,  that  $A_{2}XA_{3}$ is
  `of finite  size $(p+2)$'.

  The  operators $A_0$ and  $A_1$ 
 have two invariant subspaces,
 the  subspace  generated by the odd vectors of the basis
 $(|1\rangle,|3\rangle,...|2n+1\rangle,...)$
  and the subspace generated by $(|2\rangle,|4\rangle,...|2n\rangle,...)$.
 The action of   $A_1$ (and that  of  $A_4$)
  on both invariant subspaces is the same.
  Therefore, the product $X$   will be represented
 by  the following matrix:
\begin{eqnarray}
 X = \left( \begin{array}{cccccc}
                 \chi &0&\star&0&&\\
                 0&\chi&0&\star&. &.\\
                \star&0&\star&0&& \\
                  0&\star&0&\star&&\\
                 \star&&&.&. & \\
                  .&&&&.&.  
                   \end{array}
                    \right) \,\;
\label{prod}
\end{eqnarray}
 The symbol  $\star$ denotes unspecified
  matrix  elements. We emphasize  that the coefficients
   (1,1) and  (2,2)  of $X$ are identical.
  This coefficient is   a matrix  $\chi$
 which is a linear combination of various  products, each product
 having   $p$ factors, and  each factor being  either a $D$ or an
 $E$.

 Using the expressions  of $A_2$ and $A_3$   and the matrix  (\ref{prod})
 for  $X$,  we find
\begin{eqnarray}
   A_2 X A_3 = \left( \begin{array}{cccccc}
                 D{\chi}E - E{\chi}D  &0&0&0&&\\
                 0&0&0&0&. &.\\
                 0&0&0&0&& \\
                  0&0&0&0&&\\
                  . &&&.&. & \\
                  .&&&&.&.  
                   \end{array}
                    \right) \,\,\, ,
\label{prod2}
\end{eqnarray}
  the product $A_2 X A_3$ has only  one non-zero coefficient
 $D{\chi}E - E{\chi}D$ where  $\chi$  is
 a linear combination of  products of $p$ 
 factors  $D$  and  $E.$  Therefore, we need to show  that
 if $M$ is any   product of $p$ 
 factors  $D$  and  $E$, the matrix 
 $DME - EMD$  is finite  of size $(p+2)$ at the  most.
 This is achieved
 by induction on $p$ and by using the explicit representations
 of $D$ and $E$ given in (\ref{repr}). \hfill\break
 For $p=0$, $M = 1$ we  obtain  $DE - ED = A$ which is a   matrix
 of size 1.  \hfill\break
  Now suppose that  that our assertion is true for
 $(p-1).$ Then, let  the  matrix $M$ be a   product of $p$ factors
 $D$ and $E$;  if  $M = D M_1$ (the case $M = E M_1$ is similar),  
 we have 
 $$  DME - EMD = D (D M_1 E - E M_1 D) + (DE - ED) M_1 D \,\,\ . $$
  By  the  induction hypothesis,  $D M_1 E - E M_1 D$ is finite
 of size $(p+1)$ and multiplying it by $D$ will increase its size
  by at most  1 (one verifies  this by  using  the explicit
 representation
 given in (\ref{repr})).
 The operator $(DE - ED) M_1 D$ is equal to $ A M_1 D$ and
 is of size less than  or equal to $(p+2).$

 We have shown  that
 the factor  $ A_2 X A_3$ in (\ref{facto}) is of finite size.
 Multiplying it on the left or on the right by any of the operators
 $A_{1},A_{2},A_{3}$ and $A_{4}$ given by (\ref{mat}) does
 not alter this property since the $A_k$'s are made
 of $D$ and $E$'s  and have  only a finite number of
 non-zero  coefficients in each line and each column.
 This proves that the matrix 
  $(A_{\tau_1}...A_{\tau_L})$  has a finite trace and that
 the stationary probabilities given by
 (\ref{ansatz})  are well defined.

\subsection{Proof of the Ansatz}

 We shall use the technique  developed in \cite{sandow}
 (see for example \cite{rajew}  for details). 
 We   represent the collection  of the (unnormalized) stationary weights 
   $p({\cal C})$   as a state vector
\begin{equation}
|p\rangle={\rm Tr}( A^{ \otimes L } )
\label{state},
\end{equation}
 where  ${\otimes}$ denotes the  tensor product and   
 $A$ is a column vector, having  matrices as entries:
\begin{equation}
 A  =  \left( \begin{array}{c}
                 A_1\\
                 A_2 \\
                 A_3 \\
                 A_4 \\ 
                   \end{array}
                    \right) \,\;
\end{equation}

 This allows us  to interpret  the Markov equation (\ref{markov}) as a 
 stationary Schr\"odinger equation
 with  the non-hermitian `Hamiltonian'  $M$
\begin{equation}
  M |p\rangle =  \sum_{i=1}^{L} m_{i,i+1} |p\rangle = 0
 \label{markov2}
\end{equation}
 The matrix $m_{i,i+1}$ is local and acts only on the $i$th and
 the  $(i+1)$th element of the tensor product in (\ref{state}). 
 We show 
 that $ m_{i,i+1}[A\otimes A]$ is a divergence-like term, i.e.
  there exists a vector ${\hat A}$ defined analogously
 to $A$ 
 \begin{equation}
{\hat  A}  =  \left( \begin{array}{c}
               {\hat  A_1} \\
                 {\hat  A_2 } \\
                {\hat  A_3 } \\
                 {\hat  A_4 }  \\ 
                   \end{array}
                    \right) \,\;
\end{equation}
 such that:
\begin{equation}
 m_{i,i+1}[A\otimes A]=  A \otimes {\hat A}- {\hat A}\otimes A \,\,\,\,   .
\label{cancel}
\end{equation}
 Summation over  $i$ leads to  a global cancellation, proving thereby
 that  the Markov equation (\ref{markov2}) is satisfied.
 The proof rests upon finding four matrices 
  ${\hat A_0},{\hat A_1},{\hat A_2}$ and ${\hat A_3}$
 that satisfy equation (\ref{cancel}). In the appendix,  we write
 the  16 quadratic equations
 that  couple the $A_\kappa$'s  and the
 ${\hat A_\kappa}$'s (see equations A.1-A.3). 
    An explicit representation
 of the ${\hat A_\kappa}$'s  that solves  these equations
 is  given below 
 (here ${\bf 1}$  denotes the identity matrix)
\def\b#1{\hat #1} 
\begin{eqnarray}
 \b{A_1} &=& \left( \begin{array}{ccccc}
                 D/2+{\bf 1}&0&E/2-{\bf 1}&0&.\\
                  0&D/2+{\bf 1}&0&E/2-{\bf 1}\\
                  0&0&D/2+{\bf 1}&0&.\\
                  0&0&0&D/2+{\bf 1}&.\\
                  .&.&.&.&.
                  \end{array}
                   \hspace{0.2in} \right)\, ,
        \nonumber \\
 \b{A_2} &=& \left( \begin{array}{ccccc}
               {\bf 1}-D/2&{\bf 1}-E/2&0&0&. \\
               0&0&0&0&.  \\
               0&0&0&0&.  \\
               0&0&0&0&.  \\
              .&.&.&.&. \end{array}
                   \hspace{0.2in} \right)\, ,\nonumber  \\
 \b{A_3} &=& \left( \begin{array}{ccccc}
               E/2-{\bf 1}&0&0&0&. \\
              {\bf 1}-D/2&0&0&0&.  \\
               0&0&0&0&.  \\
               0&0&0&0&.  \\
               .&.&.&.&. \end{array}
                   \hspace{0.2in} \right)\, ,   \nonumber \\
 \b{A_4} &=& -\left( \begin{array}{ccccc}
                 E/2+{\bf 1}&0&0&0&. \\
                  0&E/2+{\bf 1}&0&0&. \\
                 D/2-{\bf 1}&0&E/2+{\bf 1}&0&.\\
                  0&D/2-{\bf 1}&0&E/2+{\bf 1}&.\\
                  .&.&.&.&.
                  \end{array}
                   \hspace{0.2in} \right)\, .
\end{eqnarray}
{\bf Remark:}
 There is one subtlety involved here.  One should not verify 
  the cancellation mechanism   only formally but also 
   make sure
 that all the traces of all the products in (\ref{cancel})
 exist. In  all   cases but  one, this follows 
 from  section 4.2.  However,
 the  case when the last factor of the trace
 is  ${A_2}{A_3}$  needs  extra care because the dynamics
 permutes these two factors. One needs  to prove  the following
 relation, where $Y$ denotes any product of the matrices
 $A_k$, $k =1,2,3,4$:
 \begin{equation}
 {\rm Tr}\{Y(A_3{\hat A_2}-{\hat A_3}A_2)\}=
 {\rm Tr}\{{\hat A_2}YA_3 - {A_2}Y{\hat A_3})\}\,.
 \end{equation}
 However, this identity can be proved  via a reasoning
   similar to that  of section 4.2.
 
\subsection{Representation--free solution}
 Consider the following choice for the $A_i$ and $\b{A_i}$
 operators:
\begin{eqnarray}
 A_1 &=& {\bf 1}\otimes D + (D-{\bf 1})^2\otimes E \, ,\\
 A_2 &=& A\otimes D + (iA(D-{\bf 1})) \otimes E \, ,\\
 A_3 &=& A\otimes E + (i(E-{\bf 1})A) \otimes D \, ,\\
 A_4 &=& {\bf 1}\otimes E + (E-{\bf 1})^2\otimes D
\end{eqnarray}
and
\begin{eqnarray}
\b{A_1} &=& {\bf 1} \otimes ({\bf 1}+D/2)+(D-{\bf 1})^2
\otimes(E/2-{\bf 1})\, ,\\
\b{A_2} &=& A \otimes ({\bf 1}-D/2) + (iA(D-{\bf 1}))
\otimes(E/2-{\bf 1})\, ,\\
\b{A_3} &=& A \otimes (E/2-{\bf 1}) + (i({\bf 1}-E)A)
\otimes(D/2-{\bf 1})\, ,\\
\b{A_4} &=& -\{{\bf 1} \otimes ({\bf 1}+E/2)+(E-{\bf 1})^2
\otimes(D/2-{\bf 1})\}\, ,
\end{eqnarray}
where again $DE=E+D$, $A=DE-ED$ (projector) and where $i=\sqrt{-1}$.
It is then possible to show that these operators
solve (A.1-A.3) and that, assuming
a third class particles at site $L$, all the traces in (\ref{ansatz}) and 
(\ref{cancel}) are real and finite. The calculation
is very similar to the one which was presented in the
preceding subsection and will therefore be omitted here.
We believe that this representation-free solution
will help to generalize our solution to the case of
a model with $N$ types of particles.

\subsection{Algebraic properties and recursion relations}

 From our solution it is straightforward to derive certain
algebraic properties of the $A_i$ operators and therefore
to find recursion relations of the relative weights
of the configurations in the system size. Some
of those relations are listed in appendix B. In fact, we found
the operators $A_i$ by solving the model for small system
sizes on the computer, guessing recursion relations
and constructing suitable operators which fulfilled
these relations. We want to remark that it seems to
us extremely unlikely that a solution could have been
constructed just by inspection of equations (A.1--A.3) of
  Appendix A. 
However, these equations turned out to be very useful for
proving that the weights given in terms
of the $A_i$'s are indeed the stationary weights.

\subsection{Finite size  cut-off}

 We have shown that the matrix Ansatz using
 the operators given in  (\ref{mat}) is
 well defined and satisfies the master equation.
 How can the representation be used for actual computations
 for systems of size $L$?
  The  proof of the finiteness of the trace
 (section 4.2) provides a method 
  to compute
 numerically the weight  of any configuration of size
 $L$ without involving infinite matrices.
 We showed that all the  matrices used  to evaluate
  the trace  are at most of size $L$. 
  Therefore,  the operators $A_\kappa$'s, $D$ and $E$
 can be restricted  to a finite size $\Lambda$,  
  with   $\Lambda$   large  enough
 to ensure that the  $L \times L$  matrices needed to 
 calculate   the weights are the same as those
 obtained by multiplying  infinite dimensional matrices.
 Such a cut-off procedure is possible
 due to the bidiagonal structure of the $A_\kappa$'s
 and  of $D$ and $E$. For example if we limit 
  $D$ and $E$ to a  finite size $N$ and
 consider  a product of $p$ such matrices, the $(N-p-1)$
 first rows and columns of the resultant  matrix 
 will be the  same as those obtained
 by taking the product of the initial infinite dimensional
 matrices.
 To  compute
 the weights of systems of size less than $L$,
 we must take  $\Lambda > 2 L$.
  The computation
 time using the matrix Ansatz grows algebraically with the system
 size whereas the increase is exponential for 
 solving the master equation.
 Thus our solution 
 allows  an exact numerical study of such systems for large sizes
 \cite{prep}.
 
 To determine the currents and the density profiles,
 one needs the normalization factor $Z$. Using
 exact results for systems of sizes up to 8, we guessed
 the following formula for $Z$ for the case when
 there is only one particle of the third type 
  ($n_3 = 1 $) and ${n_1},{n_2},{n_4} \neq 0$
 $$ Z = \frac{1}{L}{L  \choose  n_1}
{L  \choose  n_1 + n_2} { L   \choose  n_1 + n_2 + 1 }\,\,\,\, .$$
 
 \section{Conclusion}

  We have studied a generalization of the asymmetric
 exclusion process to a  system with three classes
 of particles and holes.
  This  model   can be mapped to an integrable
  two dimensional vertex model
  of equilibrium statistical physics \cite{pasquier}, but the
 Bethe Ansatz does not allow a simple
 determination  of the ground  state of this  vertex model.  
 However,  the stationary weights can be calculated via
 a matrix product Ansatz. Although analytical formulae may
 be difficult to derive (the computation of the constant $Z$
 will require  calculations similar  to those of  the diffusion
 constant of  an open system   \cite{DEMal})
 the matrix Ansatz
 enables a much faster exact numerical computation of the
 stationary state of finite size systems.

 Our main interest is  theoretical. We  wanted to know what kind
 of algebras (if any)  appear in multi-species  processes.
 Some authors  \cite{foulad,ritt,ritt2} have used generalized
 quadratic algebras to study
  systems with many species. Associativity \cite{foulad} and finite trace
 condition \cite{ritt2} for these algebras
 impose severe restrictions on the rates
 of exchange between different types of particles. The simple
 rates   we choose (\ref{rules})
 do not satisfy   these limitations. Hence
 the algebra  we have found  is not quadratic but  rather  a tensor
 product of quadratic algebras. As   emphasized in section 4.5
 and in Appendix B,
 the identities that are satisfied by the matrices $A_\kappa$
 can be cubic, quartic or of any higher order.
 We believe that  the tensor  structure we have obtained is fairly
 general. If for some special choices of the transition rates
 in (\ref{rules}), the matrices
 $D$ and $E$ can be taken to be  scalars \cite{DEHP}, 
 our matrices  (\ref{mat}) will generate  a quadratic algebra.

  There is  a  recursive
 structure when one adds new types  of particles. The
 exclusion process with only one class of particles 
 is solved by  taking  the matrices
 representing holes and particles to be both equal to 1.
 For two classes, the matrices $D$, $E$ and $A$
 are infinite dimensional matrices with
 1's   as coefficients. For the three classes  problem,
 the matrices given in (\ref{mat}) are infinite dimensional
 with $D$ and $E$ as coefficients.

 It is therefore natural to define a  generalization
 of the exclusion process for  $N$ types of particles,
 with a priority rule such that a particle of type
 $n$ can overtake a particle of type $m$ if and only if
 $ n < m $.
  This model is still integrable, and some exact
 results can be obtained via a Bethe Ansatz. Besides,
  from  numerical  solution
 of small systems  one finds  many relations between
 the rates and the matrices  representing  each type
 of particle \cite{notpers}. We hope
 that our solution will help to find a solution for this
 generalized problem. 
  
  We have studied only the totally asymmetric exclusion process,
 it may be interesting to try  to solve the partially
 asymmetric exclusion process where all  the rules in (\ref{rules}),  such 
 as `$12 \to  21 \hbox { with rate }1$',  are modified as follows
 \begin{eqnarray}
  12 &\to& 21 \hbox { with rate } p  \nonumber \\
  21 &\to& 12  \hbox { with rate } q 
 \end{eqnarray}
 with $ p + q = 1$. We believe that a suitable tensor product structure
 should allow to compute the ground state of this model. Since
 such a model could presumably display spontaneous symmetry
 breaking \cite{mukamel}, this would be of special interest.

 \hfill\break
{\bf \Large {\bf Acknowledgments}}\\

It is a pleasure to thank E. R.~Speer  for  interesting discussions,
 valuable hints and his encouragements . We also benefited from discussions
 with C. Godreche, V. Pasquier, B. Derrida, V. Rittenberg,
 J.~L.~Lebowitz and D.~Mukamel.
  K.M. acknowledges support by the Lady Davies Foundation.
  N.R. gratefully
  acknowledges a postdoctoral fellowship from the Deutsche
  Forschungsgemeinschaft and thanks Joel Lebowitz for hospitality at the
  Mathematics Department of Rutgers University and for support under NSF
  grant DMR~95--23266 and DIMACS.

\begin{appendix}
\renewcommand{\theequation}{\Alph{section}.\arabic{equation}}
\section{Explicit form of the local Markov matrix }
\setcounter{equation}{0}

 The local Markov matrix $m_{i,i+1}$
 that  describes the updating of a bond $(i,i+1)$
  is given in the canonical basis 
$(11),(12),(13),(14),(21),(22),..,(44)$ 
 by a $16\times16$ matrix:
\begin{eqnarray}\nonumber
m_{i,i+1}=   \left( \begin{array}{cccccccccccccccc}
0&0&0&0&0&0&0&0&0&0&0&0&0&0&0&0\\
0&-1&0&0&0&0&0&0&0&0&0&0&0&0&0&0\\
0&0&-1&0&0&0&0&0&0&0&0&0&0&0&0&0\\
0&0&0&-1&0&0&0&0&0&0&0&0&0&0&0&0\\
0&1&0&0&0&0&0&0&0&0&0&0&0&0&0&0\\
0&0&0&0&0&0&0&0&0&0&0&0&0&0&0&0\\
0&0&0&0&0&0&-1&0&0&0&0&0&0&0&0&0\\
0&0&0&0&0&0&0&-1&0&0&0&0&0&0&0&0\\
0&0&1&0&0&0&0&0&0&0&0&0&0&0&0&0\\
0&0&0&0&0&0&1&0&0&0&0&0&0&0&0&0\\
0&0&0&0&0&0&0&0&0&0&0&0&0&0&0&0\\
0&0&0&0&0&0&0&0&0&0&0&-1&0&0&0&0\\
0&0&0&1&0&0&0&0&0&0&0&0&0&0&0&0\\
0&0&0&0&0&0&0&1&0&0&0&0&0&0&0&0\\
0&0&0&0&0&0&0&0&0&0&0&1&0&0&0&0\\
0&0&0&0&0&0&0&0&0&0&0&0&0&0&0&0\\
\end{array}\hspace{0.2in} \right).
\label{hamiltonian}
\end{eqnarray}

 The local divergence condition (\ref{cancel}) translates
 into the following 16 coupled quadratic equations: 
\def\b#1{\hat #1} 
\begin{eqnarray}
A_i A_j &=& \b{A_i} A_j - A_i \b{A_j} \quad\quad\rm{for}\quad i<j\,,\\
A_i A_j &=& A_j \b{A_i} - \b{A_j} A_i \quad\quad\rm{for}\quad i<j\,,\\
0       &=& A_i \b{A_i} - \b{A_i} A_i \quad\quad\rm{for\,\, all}\quad i\,,
\end{eqnarray}
where $1\le i,j \le 4$.
\section{Algebraic properties and recursion relations}
\renewcommand{\theequation}{\Alph{section}.\arabic{equation}}
\setcounter{equation}{0}

The matrix algebra method is a way to encode recursion relations
 between stationary probabilities  of systems of size $L$ and of systems
 of size $(L-1)$. For some simple models, the matrices can
 be constructed using `empirical'  recursion relations
 observed on exact solutions for small systems. In our
 case, a complete set of such  relations between size
  $L$ and  size $(L-1)$ is  difficult to obtain. However
 the  matrices $A_1,A_2,A_3$ and $A_4$ given in (\ref{mat})
 satisfy a number of algebraic identities  that allow to deduce 
  some recursions  between system of different sizes. We now describe
 some of these relations that generalize the simple quadratic
 algebra (\ref{quadr}).
\hfill\break
  1.  The matrices $A_1$ and $A_4$ satisfy the algebra
 that was  found in 
  \cite{DEMal}
 and  was used to compute the diffusion constant for an open system:
 \begin{eqnarray}
   A_1 A_4^{p-1} ( A_4 A_1- A_1 A_4 )  A_1^{q-1} A_4  = \nonumber \\
  A_4^{p}A_1^{q}A_4 -  A_1 A_4^{p-1}A_1^{q}A_4
  - A_1 A_4^{p} A_1^{q-1} A_4 + A_1 A_4^{p} A_1^q
 \label{pte1}
 \end{eqnarray}
 where $p$ and $q$ are strictly positive integers.
\hfill\break
 2.  Some relations  reduce the system size and  are reminiscent of the
 $DE = D + E$ identity  in (\ref{quadr}). However,
 the following  relations are cubic and not quadratic.
\begin{eqnarray}
     A_2 A_2 A_4  &  = &  A_2 A_2 + A_2 A_4   \nonumber \\
     A_1 A_3 A_3 & = &  A_1 A_3 + A_3 A_3    \label{pt2}
\end{eqnarray}
3. Other relations are similar to the second and the
third equality in
  (\ref{quadr}):
\begin{eqnarray}
            A_2 A_2 A_3 & =  & A_2 A_3   \nonumber \\
            A_2 A_3 A_3 & =  & A_2 A_3    \nonumber \\
        A_2 A_3 A_2 A_3  &  = & A_2 A_3
 \label{pte3}
  \end{eqnarray}
 This last equality shows that
  the operator $(A_2 A_3)$ is a projector as we noted in (\ref{proj}).
\hfill\break
4. Some rules transform some particles  into others without reducing
 the size of the system:
\begin{eqnarray}
        A_1 A_2  &=& A_2 A_2 \nonumber \\
        (A_1 A_4 - A_4 A_1) A_2 &=& A_2 A_4 A_2  \nonumber \\ 
         A_3 A_4  &=& A_3 A_3 \nonumber \\ 
       A_3 (A_1 A_4 - A_4 A_1)  &=& A_3 A_1  A_3 \nonumber \\ 
       A_3 A_2 A_4 &=&  A_3 A_2 A_3 +  A_3 A_3  A_2 \nonumber  \\ 
       A_1 A_3 A_2  &=&  A_2 A_3 A_2 +  A_3 A_2 A_2
 \label{pte4}
 \end{eqnarray}

\end{appendix}

\vfill\break


\end{document}